# Film Thickness Changes in EHD Sliding Contacts Lubricated by a Fatty Alcohol

Kazuyuki YAGI[1,2]* and Philippe VERGNE[1]

[1] Laboratoire de Mécanique des Contacts et des Solides, UMR CNRS/INSA de Lyon 5514
20 avenue Albert Einstein, 69621 Villeurbanne Cedex France
[2] Graduate School of Science and Engineering, Tokyo Institute of Technology
2-12-1, Ookayama, Meguro-ku, Tokyo, 152-8552, Japan
*Corresponding author: E-mail address: kazuyuki.yagi@insa-lyon.fr

This paper describes the appearance of abnormal film thickness features formed in elastohydrodynamic contacts lubricated by a fatty alcohol. Experiments were conducted by varying the slide to roll ratio between a steel ball and a glass disk in a ball-on-disk type device. Lauric alcohol was used as lubricant and film thickness was measured in the contact area by optical interferometry. Experimental results showed that the film thickness distributions under pure rolling conditions remained classical whereas the film shape changed when the slide to roll ratio was increased. The thickness in the central contact area increased and in the same time inlet and exit film thicknesses were modified. In addition, the film shapes observed when the ball surface was moving faster than the disk one and those obtained in the opposite case were different, i.e. when opposite signs but equal absolute values of the slide to roll ratio were applied.

**Keywords:** elastohydrodynamic, film shape, sliding contact, lauric alcohol

## 1. Introduction

It is commonly accepted that the contact film thickness under elastohydrodynamic lubrication (EHL) regime depends mainly on the conditions in the inlet zone. In this zone, the fluid is dragged by the bounding surfaces towards the contact area. The viscosity of the fluid increases markedly due to the presence of a pressure gradient in the inlet. As a result, the Poiseuille flow can be ignored and thus the Couette flow is dominant within the conjunction area. Gohar et al.[1] observed experimentally that the film shape in a circular EHD contact presented a flat zone in the central area and a constriction in the exit zone in which a minimum film thickness occurred. As a consequence of numerous published studies, several equations have been proposed to predict film thickness in EHD contacts, for example those found by Hamrock and Dowson[2] in the case of point contacts. Their central film thickness equation has given a good agreement with experimental results down to a few nanometers[3].

However, several types of abnormal film shapes which were sensitive to the sliding speed have been already reported. Kaneta et al.[4,5] showed the occurrence of a conical depression (dimple) in the central part of the lubricating film in a glass-steel contact. The dimple appeared under specific operating conditions where the disk was rotating and the ball stationary. On the other hand, the film thickness shape remained classical under opposite conditions where the ball was rotating and the disk stationary. Yagi et al.[6] indicated that another type of dimple could occur in sapphire-steel point contacts under conditions where two surfaces were rotating with the same speed but in opposite directions and thus under zero entrainment speed. Guo et al.[7,8] obtained abnormal distributions in glass-steel contacts lubricated with high viscosity oils.

In the current study, the authors discuss the appearance of abnormal film thickness distributions in a glass-steel contact lubricated by lauric alcohol, i.e. a low viscosity and low pressure viscosity coefficient fluid.

## 2. Experimental Procedure

In this work, white light optical interferometry was used to measure the film thickness distribution within the contact area. This technique has been already described in detail elsewhere[9]. A circular EHD contact was obtained by applying a constant normal load between a 10 mm thick glass (BK7) disk and a 25.4 mm diameter bearing steel ball. Table 1 shows the physical properties of steel and glass. A thin chromium layer was deposited on the contacting glass disk surface to increase the interferograms contrast. A microscope was mounted above the apparatus to measure film thickness. White light was shone into the contact area through the microscope and light beams reflected at both glass/chromium and lubricant/steel interfaces were recorded by a 3CCD video camera attached to the microscope. Any correction due to the pressure influence on the lubricant refractive index has been applied because the actual pressure distribution is unknown, especially at high slide to roll ratios. Both surfaces were very smooth and rotated independently by two brushless servo motors to operate at required slide/roll ratios. The disk and the ball were surrounded by a chamber to keep the lubricant temperature constant within 0.2 °C.

Lauric alcohol used as lubricant in the tests contained 70 % wt of 1-dodecanol, 24 % wt of 1-tetradecanol and several % wt of other types of fatty alcohols. Lauric alcohol viscosity is 10 mPa.s at 40 °C and 3.7 mPa.s at 70 °C.

All experiments were carried out by applying a constant load of 30 N, which produced a maximum Hertzian pressure of 0.52 GPa and a Hertzian radius of 166 µm. The slide to roll ratio is expressed as follows:

$$S = \frac{u_b - u_d}{u_m} \quad (1)$$

where
$u_b$ and $u_d$ are respectively the ball and the disk surface velocities,
$u_m$ is the entrainment speed defined as $u_m = (u_b + u_d)/2$.

Table 1 Properties of steel and glass

|  | Unit | Steel | Glass |
| --- | --- | --- | --- |
| Young's modulus | GPa | 210 | 81 |
| Poisson's coefficient |  | 0.3 | 0.21 |
| Density | kg/m$^3$ | 7850 | 2510 |
| Thermal conductivity | W/(mK) | 46 | 1.11 |
| Specific heat capacity | J/(kgK) | 470 | 840 |

## 3. Results

Figure 1 shows the interferograms obtained with an entrainment speed $u_m$ of 1.8 m/s, a temperature $T$ of 40 °C and for various positive slide to roll ratios. $S > 0$ means that the ball surface is moved faster than the disk surface.

Under pure rolling conditions, the shape of the lubricating film formed within the contact area presents a flat zone in the central area and a horse-shaped constriction in the exit area. The shape of the film thickness is similar to those reported in the pioneered paper published by Gohar et al[1]. The central film thickness is close to 120 nm. The pressure-viscosity coefficient was deduced from a comparison between predicted values (Hamrock and Dowson equation[2]) and experimental results and its value at 40°C is 10.1 GPa$^{-1}$.

Under rolling and sliding conditions, the film thickness distribution changes. From pure rolling to a slide to roll ratio of 135 %, the color in the central part of the interferograms varies gradually while keeping an almost similar shape. The slight color change indicates a moderate increase of the central film thickness. When the slide to roll ratio exceeds 135 %, the film shape turns into a quite different aspect. Firstly, the thicker feature moves towards the contact inlet. Secondly, the grey-black area expands around the downstream side that means the thinnest film thickness zone spreads in this area.

Figure 2 depicts the interferograms obtained for negative slide to roll ratios, i.e. when the disk surface was moved faster than the ball surface. Under these conditions, the film thickness shape variations remain quite similar to those observed when positive slide to roll ratios are imposed. However, the colors found in the two sets of interferograms are different. The color difference shows that the film thicknesses do not coincide even if the same absolute values of the slide to roll ratio are applied.

Figure 3 presents the film thickness profiles measured on the contact center line along the sliding direction for various slide to roll ratios. The film thickness in the conjunction area increases gradually together with a more pronounced augmentation in the profile central part when the slide to roll ratio increases from pure rolling to $S = 135$ %. At $S = 165$ %, the film thickness reaches a maximum of about 350 nm at $x = -105$ µm, near the leading side of the contact area. At $S = 180$ %, the oil film collapses markedly in the exit region and the thickness falls down to about 90 nm while in the leading zone the oil film thickness increases and is much more thicker than the one found under pure rolling conditions.

For applied negative slide to roll ratios, the film thickness variations are qualitatively similar. However, the thickening observed in the central contact area is more important than the one observed when positive slide to roll ratios were imposed. A maximum film thickness of about 440 nm is found at $S = -165$ %.

## 4. Discussion

The film distributions shown in Figures 1 and 2 are really different from the observations made by Gohar et al.[1] that are known to be representative of film thickness shape occurring in EHD circular contacts. In our case, the thickness distribution varies with the slide to roll ratio. Kaneta et al.[4,5] indicated that a conical depression of the film thickness could occur in glass-steel contacts. The bounding surface materials that these authors employed were the same than those used in this study and the dimple formation that they described was also sensitive to the slide to roll ratio. The dimples observed by Kaneta et al.[4,5] appeared only when the disk surface was sliding and the ball surface stationary while the thickness shape remained classical except a slight thickness increase just behind the constriction region when the ball surface was moving and the disk stationary. In contrast, the abnormal film shapes reported in this study appear under both positive and negative slide to roll ratios.

On the other hand, the film thickness variations presented in this study are similar to those obtained by Guo et al.[7,8], who showed a film thickness increase in the central area under moderate slide to roll ratios. With a further slide to roll ratio increase the thicker feature moved towards the inlet region. A thicker film in the conjunction appeared under both positive and negative slide to roll ratios as in our results presented in Figures 1 and 2. In their experiments, Guo et al.[7,8] used several types of polybutene, whose viscosity was several orders of magnitude higher than lauric alcohol viscosity and the sliding speed was low compared with those imposed in this study. The low sliding speed allowed to ignore temperature rise in the lubricating film, while in the present study thermal effects may influence the formation of the oil film when the sliding speed is high. Therefore, it would be necessary in our case to estimate the temperature rise in the conjunction area to investigate the mechanisms of the quite unexpected film shape formation.

## 5. Conclusion

In this study, film thickness in a lubricated circular contact was measured by white light optical interferometry at various slide to roll ratios. Lauric alcohol was used as lubricant: it has a low viscosity and low pressure-viscosity coefficient compared to oils that were used in studies where abnormal EHD film shapes or dimples were observed. Above results and discussion are summarized as follows.
1) The film thickness distribution varies when the slide to the roll ratio is increased.
2) From pure rolling to an absolute value of the slide to roll ratio of 135 % the film thickness increases. Beyond 135 %, the thicker feature moves from the central part of the contact towards the inlet zone while at the exit zone a thinner area expands.
3) Film thicknesses obtained at the same absolute values but at positive and negative slide to roll ratios are different.

## 6. Acknowledgements

The first author has been financially supported by a Research Fellowship of the Japan Society for the Promotion of Science (JSPS) for Young Scientists. The authors wish to thank Professor Nakahara of the Tokyo Institute of Technology in Japan for his valuable comments.

## 7. References

[1] Gohar, R.; Cameron, A. Nature, 1963, 200, 458.
[2] Hamrock, B. J.; Dowson, D. Trans. ASME. J. Lub. Tech., 1977, 99, 246-276.
[3] Glovnea, R. P.; Olver, A. V.; Spikes, H. A. Trib. Trans., 2005, 48, 328-335.
[4] Kaneta, M.; Nishikawa, H.; Kameishi, K.; Sakai, T.; Ohno, N. Trans. ASME. J. Trib., 1992, 114, 75-80.
[5] Kaneta, M.; Nishikawa, H.; Kanada, T.; Matsuda, K. Trans. ASME. J. Trib., 1996, 118, 886-892.
[6] Yagi, K.; Kyogoku, K.; Nakahara, T. Trans. ASME. J. Trib., 2005, 127, 658-665.
[7] Guo, F.; Wong, P. L. Trans. ASME. J. Trib., 2005, 127, 425-434.
[8] Guo, F.; Wong, P. L. In Proceeding of IUTAM Symposium on Elastohydrodynamics and Microelastohydrodynamics, Eds by Snidlle R. W. and Evans H. P., Springer, The Netherlands 2006 285-296.
[9] Hartl, M.; Krupka, I.; Poliscuk, R.; Liska, M.; Molimard, J.; Querry, M.; Vergne, P. Trib. Trans., 2001, 44, 270-276.

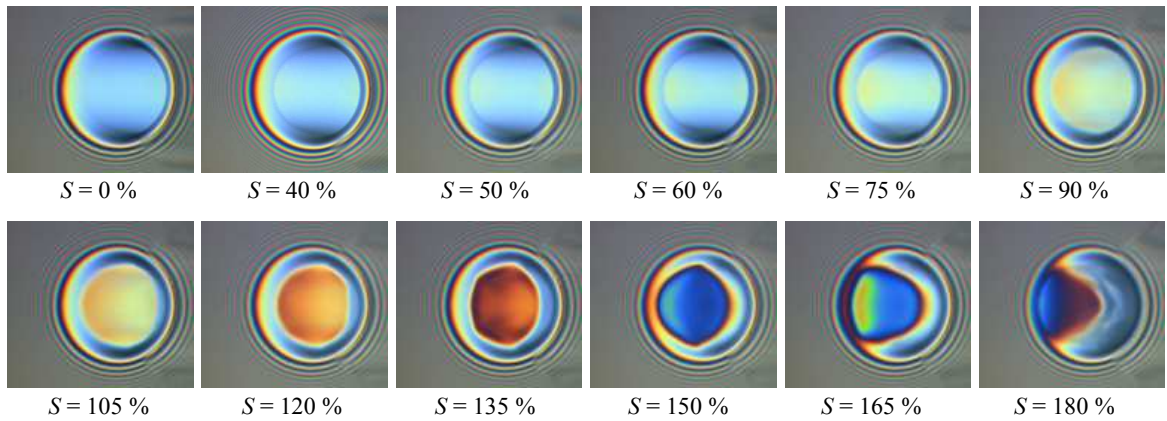

Fig.1 Interferograms obtained for various positive slip ratios ($u_m$ = 1.8 m/s, $p_{max}$ = 0.52 GPa, $T$ = 40 °C, lauric alcohol)

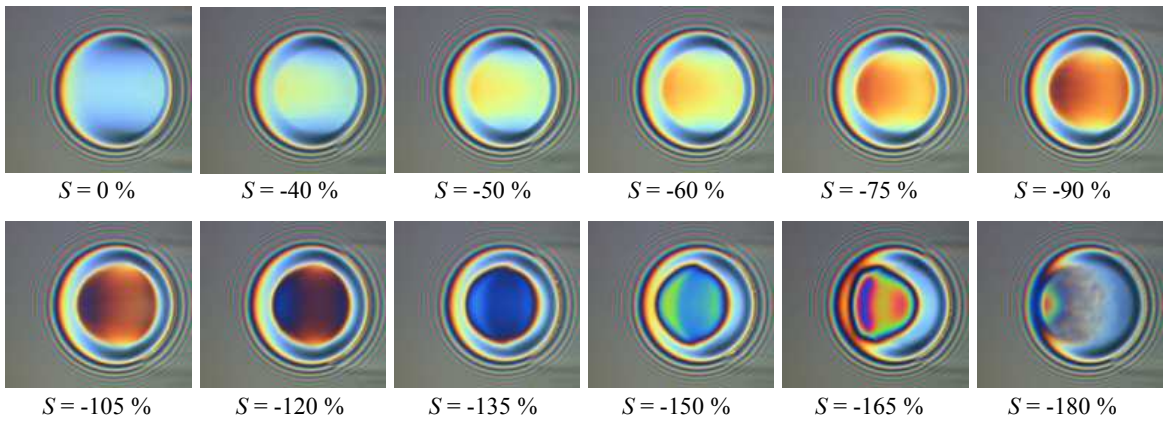

Fig.2 Interferograms obtained for various negative slip ratios ($u_m$ = 1.8 m/s, $p_{max}$ = 0.52 GPa, $T$ = 40 °C, lauric alcohol)

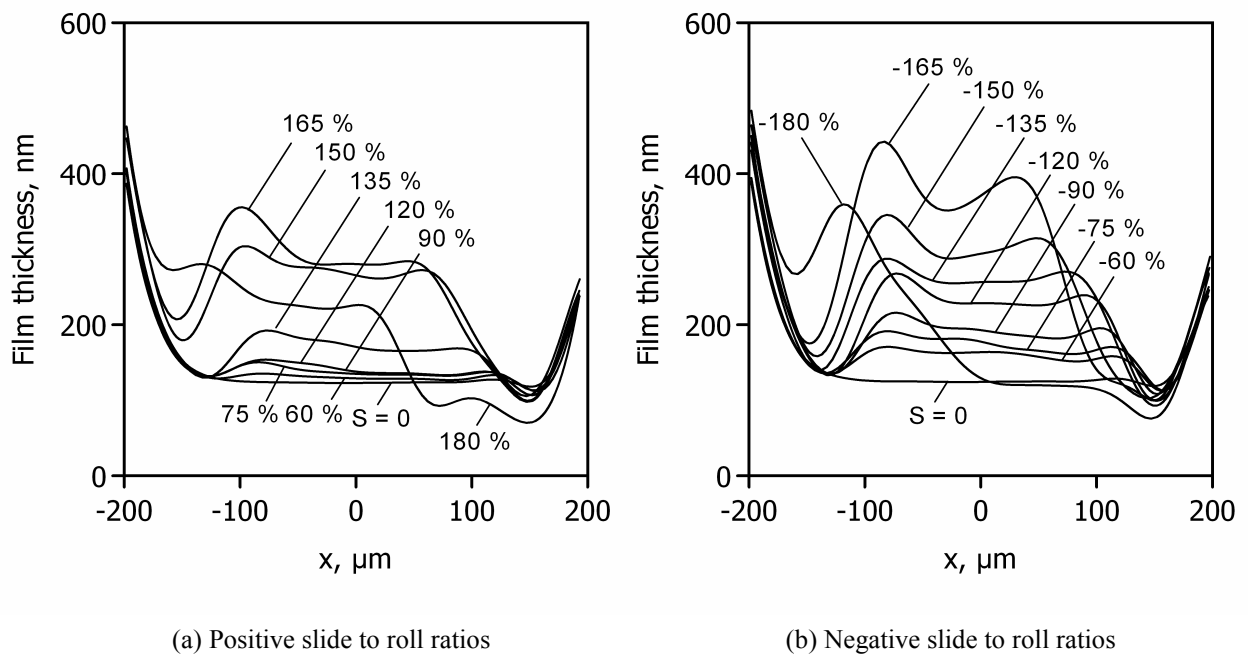

(a) Positive slide to roll ratios  (b) Negative slide to roll ratios

Fig.3 Profiles of film thickness on center line along sliding direction for various slip ratios ($u_m$ = 1.8 m/s, $p_{max}$ = 0.52 GPa, $T$ = 40 °C, lauric alcohol)